\documentstyle[preprint,aps,prl,psfig]{revtex}
\begin{document}
\newcommand{\sinf}{\raisebox{-.7ex}{$\stackrel{<}{\sim}$}}
\newcommand{\ssup}{\raisebox{-.7ex}{$\stackrel{>}{\sim}$}}

\ifx\undefined\psfig\def\psfig#1{    }\else\fi
\ifpreprintsty\else
\twocolumn[\hsize\textwidth
\columnwidth\hsize\csname@twocolumnfalse\endcsname       \fi    \draft
\preprint{  } \title  {Intrinsic electric field effects on few-particle 
interactions in coupled GaN quantum dots}
\author  { S. De Rinaldis$^{1,2,3,*}$, I. D'Amico$^{1,4}$ and 
F. Rossi$^{1,4,5}$}
\address{ $^1$INFM- Istituto Nazionale per la Fisica della 
Materia;\\$^2$NNL- National Nanotechnology Laboratory of INFM, via
per
Arnesano, 73100, Lecce, Italy;\\$^3$ISUFI- Istituto Superiore
Universitario
per la Formazione Interdisciplinare, via per
Arnesano, 73100, Lecce, Italy;\\$^4$Institute for Scientific 
Interchange (ISI), Villa Gualino, Viale Settimio Severo 65, I-10133 
Torino, Italy;\\$^5$Dipartimento di Fisica, Politecnico di Torino, 
Corso Duca degli Abruzzi 24, I-10129 Torino, Italy}
\date{\today} \maketitle

\begin{abstract}

We study the multi-exciton optical spectrum of vertically
coupled GaN/AlN quantum dots with a realistic three-dimensional 
direct-diagonalization approach
for the description of few-particle Coulomb-correlated states. 
We present a detailed  analysis of  the fundamental properties of
few-particle/exciton interactions peculiar of nitride materials. 
The giant intrinsic electric fields and the high electron/hole 
effective masses give rise to different effects compared to GaAs-based 
quantum dots: intrinsic exciton-exciton coupling, non-molecular 
character of coupled dot exciton wavefunction, strong dependence of the 
oscillator strength on the dot height, large ground state energy shift
for dots separated by different barriers. Some of these effects make GaN/AlN
quantum dots interesting candidates in quantum information processing.

{$^*) Electronic$ mail: srinaldi@chem.utoronto.ca}\newline 
Present address: Chemical Physics Theory Group, Department of Chemistry
University of Toronto, 80 St. George Street, Toronto, Ontatio M5S 3H6, 
Canada\newline 
\end{abstract}

\ifpreprintsty\else\vskip1pc]\fi \narrowtext

\section {Introduction}

In the last 
few years semiconductor quantum dots (QDs) has been an area of  intensive research  
in condensed matter physics \cite{dot}. 
The possibility of tailoring some of their structural
parameters represents a 
fundamental tool in order to study basic physics and to achieve 
technological applications \cite{Saito}, \cite{Finley}, \cite{Bruchez}.
The strength of the lateral or vertical 
 coupling between  two quantum dots
represents a further degree of freedom for engineering optical and 
electrical properties  and for studying  few-particle phenomena 
\cite{ortner}, \cite{shtrichman};   such coupled nanostructures are also
 known as ``artificial'' or ``macro'' 
molecules. 
 
The growth of GaN/AlN quantum dots  \cite{arakawa} and the demonstration of 
their vertical and lateral alignment  \cite{brault} has been recently 
achieved.
While GaAs-based quantum dots have been widely studied both theoretically and 
experimentally \cite{Rinaldi}, \cite{Biolatti1}, \cite{Biolatti2},  GaN 
quantum dots are 
becoming a subject of 
increasing interest for their potential technological applications, such as
 quantum information 
processing \cite{DeRinaldis}, single electron read-out devices
  \cite{damicorossi} or spintronics   \cite{awschalom}.

Our analysis is focused on few-particle effects in 
nitride dots.  We address some 
distinguished 
few-particle phenomena typical of nitride QDs and mainly stemming from the built-in giant electric field which characterizes such nanostructures. 
We analyze the behavior of the
intrinsic exciton-exciton dipole coupling as a function of the various structure parameters. In addition our calculations show that the ground state excitonic
transition of two identical GaN 
dots is characterized by
a strong blueshift when 
their relative distance is decreased.
 The corresponding  ground-state excitonic wavefunction preserves its
 atomic-like behavior,
in contrast to what happens in
 GaAs-based couple QDs, where a redshift of the ground-state and 
a splitting in
bonding and anti-bonding molecular-like states is observed\cite{fafard}. 
We stress that our analysis is also relevant for 
the experimental realization of the quantum information processing 
strategy 
proposed in \onlinecite{DeRinaldis}, in which a large biexcitonic shift is 
necessary 
for energy selective addressing of the different few-particle excitation 
with 
fs/ps laser pulses.

The paper is organized as follows. In section II we compare the characteristics of GaAs and GaN-based heterostructures, in sec. III we 
develop our theoretical 
approach; in section IV we introduce the investigated
system; in section 
V we focus on exciton-exciton interactions and on 
 the  oscillator strength 
of the transitions.
In section VI we study two  coupled identical GaN dots 
 and we compare GaAs and GaN macro molecules investigating the 
role of the different parameters involved.
In section VII the main results are summarized and some 
conclusions are drawn.

\section {GaN versus GaAs-based quantum dots}
Let us first of all compare the general characteristics of GaN-based quantum dots
with the more familiar  GaAs-based ones.
The GaAS compounds have a zincblende structure characterized by a face-centered cubic cell. This structure does not present spontaneous polarization, and, in the absence of applied electric fields, the dipole moment of correlated electron-hole excitations will be negligible and so will be the interaction between excitons created in different QDs.
For application purposes however, it would be desirable to have a controllable source of interaction between different QDs, since the latter 
can be naturally thought as the different units of a device.
A possibility for creating such an interaction has been envised and studied in \onlinecite{Biolatti1,Biolatti2}. 
The underlying idea is to create interacting dipoles, i.e. to
 separate electron and hole excitonic components
 by means of an in-plane externally
 applied (static) electric field.

The great advantage of 
III-V nitride compounds, as GaN, is that they may present {\it intrinsic} electric fields, and, as a consequence, they automatically
 have {\it built-in} such an interaction.
III-V nitride compounds may present in fact a wurzite type of structure
 based on a hexagonal cell, which is compatible with spontaneous bulk polarization, i.e. they present a nonvanishing dipole per unit volume. 
In the heterostructures we consider, this polarization is accumulated at the GaN/AlGaN interfaces due to the asymmetry  of
GaN and AlN unit cells.

 In the case of QDs  a strong  strain-induced piezoelectric field along the (0001) direction must be added to this effect.
This is considerably stronger than in GaAs-based nanostructures because 
the piezoelectric 
constants in nitrides are orders of magnitude greater than in other III-V 
compounds. 
 The sum of the two contributions results in a strong built-in electric field of the order of few 
MV/cm which is oriented along the growth direction, and  has opposite sign inside and outside the dot.
GaN-based QD are then characterized by strong intrinsic electric fields which automatically creates intrinsic dipoles out of each and every electron-hole correlated pair excitation. 

Let us consider two vertically stacked QDs coupled by the exciton-exciton
interaction just described. 
In GaAs the particle distributions corresponding to the  biexcitonic ground state generated by laser pulses having the same polarization,
 will approximately correspond  to two parallel dipoles (one for every
 QD) aligned along the (in-plane) field direction\cite{Biolatti2}.
 In this way excitons in two vertically stacked QDs will have a positive Coulomb interaction.

If the corresponding GaN based structure instead, such a state is formed
by approximately two
dipoles but stacked in the growth direction. Their interaction energy will  then be negative.
This interaction energy is the so called biexcitonic shift, i.e. the difference in energy between the transition energy corresponding to the creation of a certain exciton and the one corresponding to the same transition but in the presence of another exciton.

In GaAs based structures the biexcitonic shift can be tuned by  engineering dots parameters as height and base and by modifying the value of the external applied field.
In the case of GaN based  the strength of the built in field is instead determined by the structure parameters.
In both materials and for experimentally reasonable parameters, it is possible to achieve (at least theoretically ) biexcitonic shifts of the order of few meV,  which are consistent with fast, sub-picosecond,
 exciton dynamics.

From the previous analysis it is clear that both GaAs and GaN
structures may present a tunable exciton-exciton coupling which can be used for different applications and in particular to perform two-qubit conditional operations\cite{Biolatti1,Biolatti2,DeRinaldis}.
One drawback of applying an external electric field (as needed in GaAs based structures) is that a too strong electric field might ionize
 the trapped charges.
An externally applied field also means more complex overhead 
circuits.

Due to the strength of the built-in electric field,
 the difference in energy between the two lowest excitonic transitions is one order of magnitude bigger in GaN-based structures than in the GaAs-based ones. Additionally in the GaN-based structures
the oscillator strength associated to ground state excitons in QDs of  different heights 
are different because the QDs height roughly 
determines the dipole length and this strongly influence
 electron and hole wavefunction
 overlap.

\section {Theoretical model}
                                                                                
The theoretical approach employed to study the optical response of our GaN
nanostructure is
a  fully three-dimensional
exact-diagonalization scheme, as described  in \onlinecite{Biolatti2}.
The confinement potential of GaN  "macromolecule" is modeled
as parabolic in
the $x-y$ plane and as a sequence of
triangular-like potential wells along the growth ($z$) direction (see 
Fig.1).

The physical system under investigation is a gas of electron-hole pairs
confined in a semiconductor quantum dot or in two vertically coupled 
quantum dots.
The Hamiltonian is the sum of three terms:
\begin{equation}
H=H^c+H^{cc}+H'
\end{equation}

The term $H^c$ is a sum of single-particle Hamiltonians which
 describes a gas of noninteracting carriers,  electrons and
holes, and includes the QD quasi-0D confining potential. 
The term $H^{cc}$  describes the
correlation of the carriers via the two-body Coulomb interaction. The term 
$H'$ is a light-matter
interaction Hamiltonian, which accounts for the laser light absorption in
a quantum dot.
We consider the multi-exciton optical spectra, i.e., the absorption probability
corresponding to the generic $N \to N'$ transition, where $N$ is the 
exciton number, and in particular we evaluate\onlinecite{Biolatti2}.
 the excitonic ($0\to 1$) and biexcitonic ($1\to 2$) optical spectra.
The biexcitonic $(1 \to 2)$ optical spectrum
describes the creation of a second electron-hole pair in the presence
of a previously generated exciton. Here, we shall consider
parallel-spin configurations only.
 In the next three subsection we will describe the three 
terms of the Hamiltonian.

\subsection {Single particle description}
We will work in the  usual effective-mass \cite{yucardona} and  
envelope-function approximation \cite{bastard}.
Within such approximation scheme,
the {\it non-interacting} 
electron and holes wavefunctions are
described by the following
Schrodinger equation:

\begin{equation}
-\frac {{\hbar}^2{\nabla}^2} {2m^*_{e/h}}{\psi}_{e/h} 
+V^{e/h}{\psi}_{e/h}=E_{e/h}{\psi}_{e/h}
\label{region3}
\end{equation}

where e/h describes the set of quantum numbers for electrons (e) and holes
(h); ${\psi}$ is the
envelope function of the quantum state; $m^*_{e/h}$ is the bulk effective
mass for electrons (e)
and  holes (h); $E_{e/h}$ are the energy levels; $V^{e/h}$ is the
three-dimensional
confinement potential.
Since we are interested only in the lowest energy levels, we can
approximate the confinement
potential as the sum of two potential profiles acting in the  parallel and
perpendicular
direction to the growth plane. Moreover, a 2D parabolic potential in the
x-y plane has been
proven to reproduce experimental data \cite{Rinaldi}. Therefore in our
model we consider a 2D
parabolic potential, whose analytic solutions are known, and we solve
numerically the Schrodinger
equation along the growth direction using a
plane-wave expansion technique \cite{rossiwires}. We expand the unknown 
envelope function in a plane-wave basis:
                                                                                
\begin{equation}
{\psi}_{e/h}=\frac {1} {\sqrt{L}}{\sum}_kb_ke^{ikr}
\label{region2}
\end{equation}

We use periodic boundary conditions with a box of length L;
$k=n(2{\pi}/L)$ are the reciprocal
lattice vectors. By substituting the plane-wave expansion (\ref{region2}) 
in the Schrodinger equation (\ref{region3})
we obtain an eigenvalue equation:

\begin{equation}
{\sum}_k(H_{kk'}-E{\delta}_{kk'})b_{k'}=0
\end{equation}

where $H_{kk'}$ are the matrix elements of the single-particle Hamiltonian
in a plane-wave basis.
A direct diagonalization can be done with standard commercial packages. 
The electron and hole states expressed in terms of  
creation/destruction operators are:

\begin{eqnarray}
\nonumber|e{\rangle}=c^{\dagger}_e|0{\rangle}\\
|h{\rangle}=d^{\dagger}_h|0{\rangle}
\end{eqnarray}

where $|0{\rangle}$ is the vacuum state.
The single-particle Hamiltonian of the system can be rewritten as
                                                                                
\begin{equation}
H^c=H^e+H^h={\sum}_eE_ec^{\dagger}_ec_e+
{\sum}_hE_hd^{\dagger}_hd_h
\end{equation}

\subsection {Coulomb correlations}
                                                                                
The single-particle eigenfunctions and the corresponding eigenvectors are
the input for the
calculation of Coulomb interactions between the carriers.
In writing $H^{cc}$ we neglect
Auger processes, that
take place only at high particle densities and energy far from the bottom
of the band. Only
processes conserving the number of carriers are considered:

\begin{eqnarray}
\nonumber H^{cc}=H^{ee}+H^{hh}+H^{eh}=\\
\nonumber\frac{1}{2}{\sum}_{e_1e_2e_3e_4}V_{e_1e_2e_3e_4}
c^{\dagger}_{e_1}c^{\dagger}_{e_2}c_{e_3}c_{e_4}+\\
\nonumber\frac{1}{2}{\sum}_{h_1h_2h_3h_4}V_{h_1h_2h_3h_4}
d^{\dagger}_{h_1}d^{\dagger}_{h_2}d_{h_3}d_{h_4}+\\
-{\sum}_{e_1e_2h_1h_2}V_{e_1e_2h_1h_2}c^{\dagger}_{e_1}
d^{\dagger}_{h_1}d_{h_2}c_{e_2}
\end{eqnarray}

The physical interpretation of the terms in $H^{cc}$ is respectively,
electron-electron and
hole-hole repulsive interaction ($H^{ee}$ and $H^{hh}$ ) and electron-hole
attractive
interaction ($H^{eh}$), while V
indicates the the Coulomb potential matrix for a generic
two-particle transition.
After having obtained by direct diagonalization the 3D single-particle
electron and hole
wavefunctions, we calculate the matrix elements of the complete
many-body Hamiltonian ($H^c+H^{cc}$ )
in the basis of
products of electron and hole eigenstates of the
single-particle Hamiltonian.
We consider many-body state with the same number of electron and holes,
that are in general
denoted as exciton (N=1), biexciton (N=2), etc. 
 By direct diagonalization we obtain the energies and
wavefunctions, that will be
expressed as a linear combination of products of single-particle states.\newline

\subsection {Matter-light interaction}

The light-matter interaction Hamiltonian is:
                                                                                
\begin{equation}
H'=-E(t)({\sum}_{eh}{\mu}^*_{eh}c^{\dagger}_ed^{\dagger}_h+
{\sum}_{eh}{\mu}_{eh}c_ed_h)
\end{equation}

E(t) is the electromagnetic field of the laser and ${\mu}_{eh}$  is the
dipole matrix element
for the e-h transition that can be factorized as a "bulk" and a
wavefunction dependent part:

\begin{equation}
{\mu}_{eh}={\mu}_{bulk}{\int}{\psi}^*_e(r){\psi}_h(r)dr
\end{equation}

The matrix elements of the light-matter Hamiltonian are different from
zero only for the
transitions
$N{\rightarrow}(N+1)$ and $(N+1){\rightarrow}N$  that correspond
 to the absorption and the
emission of a photon, respectively (N is the number of
electron-hole pairs,
that is conserved  by $H^c+H^{cc}$).
Following the Fermi's golden rule we can calculate the absorption
probability for the
$N{\rightarrow}(N+1)$
transition:
                                                                                
\begin{equation}
P({\lambda}_N{\rightarrow}{\lambda}_{N+1})_E=\frac{2{\pi}}{\hbar}
|H'({\lambda}_N{\lambda}_{N+1})|^2
{\delta}(E({\lambda}_{N+1})-E({\lambda}_N)-E)
\end{equation}
where the state $|\lambda_N\rangle$ corresponds to N Coulomb-correlated
 electron-hole pairs.

\section {Coupled GaN dots}

We study the multi-exciton optical response of GaN/AlN QDs  
in the range of parameters experimentally realized in \onlinecite{Arley}: 
the dot height is varied between $2$ to $4$\,nm and the 
base 
diameter is varied from
$10$ to $17$\,nm, proportionally 
to the height (according to experimental data in \onlinecite{Arley}). 
The material in the dot is assumed to have a 
constant composition 
of GaN, while in the barrier is pure AlN, thus neglecting intermixing. The 
valence band discontinuity 
is set, according
to experimental values, to 0.5 eV, the conduction band to 2.0 eV 
\onlinecite{king}. The typical system considered is composed by two 
QDs stacked along the growth $z$-axis (see
Fig.~1); in the in-plane directions the confined potential is 
assumed to be parabolic. 
In Table I we compare the different parameters used in this paper for GaN  
and GaAs quantum dots. GaN has higher electron/hole masses and 
conduction/valence band discontinuities.
The main feature of wurzite compared to zincoblende GaN heterostructures 
is the strong
built-in electric field.  
 The strength of the
intrinsic field  is of the same order of magnitude inside and outside the dot 
(MV/cm), 
but it is
opposite in sign, antiparallel to the growth direction inside the dot. 
The built-in electric field in GaN QDs and AlN barriers is calculated
according to\cite{Cingolani}:

\begin{equation}
F_{d} ={L_{br}(P_{tot}^{br}-P_{tot}^{d})\over\epsilon_{0}(L_{d}\epsilon_{br}+
L_{br}\epsilon_{d})}\,
\label{region1}
\end{equation}

where $ \epsilon_{br,(d)}$  is the relative dielectric constant
of the barrier (of the
quantum dot), $P_{tot}^{br,(d)}$  is the total polarization
of the barrier (of the
quantum dot), and
$L_{br,(d)}$ is the
width of the
barrier (the height of the dot).
The value of the field in the barrier $F_{br}$   is obtained by
exchanging the
indices {\it br} and {\it d}.
Equation (\ref{region1}) is derived for an alternating sequence of
quantum wells and barriers, but it is a good
approximation also in the case of   an array of
similar QDs  in the growth (z) direction.
In our approach the in-plane components
 of the  built-in electric field 
are in fact ``absorbed'' in the  strongly  
confining bidimensional
parabolic potential which in addition
 preserves the spherical symmetry 
of
the ground state\cite{Andreev}. Our modeling
is supported by the agreement  with the
experimental findings in \onlinecite{Andreev}.
The polarization $P_{tot}^{br,(d)}$
is the sum of the spontaneous polarization charge that
accumulates at GaN/AlN interfaces and the piezoelectric one, 
$P_{tot}^{br/d}=
P_{piezo}^{br/d}+P_{sp}^{br/d}$. The piezoelectric charge is induced by the 
lattice mismatch
and by the thermal strain 
($P_{piezo}^{br/d}=P_{ms}^{br/d}+P_{ts}^{br/d}$).
All the material parameters are the one
used in ref \cite{Cingolani}, except for $P_{sp}^{br}$ and 
${\sigma}_{||}^{br/d}$ 
which depend on the Aluminum concentration in the barrier:
in the quantum 
wells 
considered in \cite{Cingolani} such concentration 
is Al=0.15 while for GaN/AlN quantum dots 
is Al=1.
In particular ${\sigma}_{||}^{br/d}=2.48\%$, therefore in GaN-based
 dots 
spontaneous
polarization and piezoelectric fields have similar values while in 
$Al_{0.15}Ga_{0.85}N/GaN$ quantum well the piezoelectric field contributes 
only 10\%
of the total value. The other parameters necessary to calculate the 
electric field are:
$P_{ms}^{br/d}=-2(e_{33}C_{13}/C_{33}-e_{31}){\sigma}_{||}^{br/d}$,
$P_{ts}^{d}=-3.2*10^{-4}C/m^2$, $P_{ts}^{br}=0C/m^2$, $P_{sp}^{br}=-0.081C/m^2$
 and
$P_{sp}^{d}=-0.029C/m^2$.

The giant internal field strongly modifies the conduction and valence
bands along the growth direction and  causes the separation of electrons 
and
holes, driving the first one toward the QD top and the latter
toward its bottom. This corresponds to
the creation of intrinsic dipoles.  
In this context, the charge distribution of two 
vertically coupled GaN dots
occupied by one exciton each can be described as
 two dipoles aligned along the
growth direction. This is exemplified in   
Fig.1, where  we plot the electron and hole single-particle 
distributions
corresponding to the lowest biexcitonic state (with parallel-spin 
excitons)
for one of the GaN/AlN  quantum dot nanostructure considered.
Nearby quantum dots are then  coupled by the 
corresponding exciton-exciton (dipole-dipole) interaction.

\section {Biexcitonic shift and oscillator strength}
The energy renormalization of the excitonic transition in the presence of
 another 
exciton  is known as biexcitonic shift.
A biexcitonic shift of the order of few meV in two coupled GaN dots is the 
prerequisite for
the implementation of conditional quantum dynamics in the quantum information/computation scheme proposed 
in 
\onlinecite{DeRinaldis}.
In the aforementioned scheme excitons in different QDs are manipulated by 
energy-selective addressing;
additionally, due to the strain field, two stacked QDs are in general not identical,
 therefore we set the difference between the 
well widths of 
two stacked QDs to be 8\% \cite{madhukar}.
This  allows for energy-selective generation of ground-state excitons in 
neighboring QDs. 
For the range of parameters considered, the barrier width is such to 
prevent single-particle 
tunneling and to allow at the same time significant dipole-dipole Coulomb 
coupling.
In contrast to GaAs quantum dots, even with a 2 nm barrier the 
single-particle tunneling is 
negligible on the nanosecond time scale (i.e. on a time-scale comparable to the excitonic lifetime).
This is due to the higher valence/conduction band gaps and electron/hole 
effective masses.

The theoretical approach employed to study the optical response of our GaN 
nanostructure is 
a  fully three-dimensional
exact-diagonalization scheme, as described  in Sec.III. 
The confinement potential of GaN  "macromolecule" is modeled 
as parabolic in 
the $x-y$ plane and as a sequence of
triangular-like potential wells along the growth ($z$) direction (see Fig.1).
We evaluate the 
excitonic ($0
\to 1$) and biexcitonic ($1
\to 2$) optical spectra.
The biexcitonic $(1 \to 2)$ optical
spectrum
describes the creation of a second electron-hole pair in the presence
of a previously generated exciton. Here, we shall consider
parallel-spin configurations only.
For all the structures considered, the two lowest optical
transitions are, respectively, direct ground-state
excitons in dot $a$ and $b$ (see Fig.~1).
In Fig.~2 we show an example of such an absorption spectrum for the 
parameters of Fig.1.
In such example a negative exciton-exciton coupling $\Delta\varepsilon=5.7$
 meV is the signature of
the creation of vertically
aligned dipoles forming the biexcitonic ground state.

Let us focus on the biexcitonic shift defined in the present case as
the energy difference between the ground-state biexcitonic 
transition 
(given a ground-state exciton in dot $a$)
and the ground-state excitonic transition of dot $b$.
This quantity 
 provides  the essential coupling for realizing conditional gates
in exciton-based all-optical quantum information 
schemes\cite{Biolatti1,DeRinaldis}.

The biexcitonic shift can be engineered by varying the coupled GaN dots 
parameters 
with self-assembled growth. We analyze how it depends on the height and base 
diameter of the dot
and barrier between the dots. We  also study
 the corresponding variation of the 
oscillator strength 
of the transition. 
Figure 3 shows how  the biexcitonic shift
depends   on the dimensions (height $L_d$ and base diameter $D$)
of the dot, for a  barrier width that is
equal to $2.5$\,nm.
We have considered three cross-sections of the space parameters, the first (solid line, A) keeping the QD height fixed ($L_d=2.5$ nm), the second (dashed line, B) fixing the base 
diameter to $D=10$ nm (small dots), and the third (dashed-dot line, C) varying the base diameter proportionally to the height, according
to the experimental relation $D = 3.5L_d+3$\,nm\cite{Arley}.
The height and base diameter values plotted corresponds to the smaller dot. 

The excitonic dipole length is roughly
proportional to the height of the dot because
of the MV/cm built-in electric field; therefore 
the
dependence of the exciton-exciton interaction on the  QD height 
is the strongest (curve B). 

By looking at curve A, instead, we notice that
 the spreading of the wavefunction related to a wider dot basis,
 negatively affects the biexcitonic 
shift: in fact this decreases from  5.1 to
4.3 meV, as the base diameter goes from 10 to 17 nm\cite{Arley}.  
Exciton-exciton interaction  is favored by  "localized" states, 
virtually achieving a maximum in an idealized ``point-like'' particle case.

Curve C presents the experimentally most common case 
in which base and height of the dots change simultaneously.
  For realistic parameters the biexcitonic shift can be up to 20\% smaller
than in curve B, 
where the wavefunction is more localized, being the diameter  fixed to the 
value of the
"small" dots. In such case  it is possible to achieve
biexcitonic shifts up to $9.1$\,meV.

Our results show that the 
best strategy to achieve large biexcitonic 
shift is to grow
"high" and "small diameter"  dots. Unfortunately 
 the oscillator strength (OS) of
the ground-state transition strongly decreases with the height of the dot,
since it is proportional to the overlap of electron and hole wave 
functions. Since the coupling of the laser field to the considered transition is directly proportional to this quantity, 
in view of an all-optical manipulation of correlated electron-hole pair excitations, it is of 
utmost importance to study the dependence of the OS on the various parameters.
In
Fig. 4   the OS corresponding to the excitonic 
ground state
 of dot $b$ is plotted for the same parameters of Fig. 3 (same labeling of the curves). Curve B (fixed base) shows that the OS
decreases super-exponentially   with the height of the dot.
It changes over three order of magnitude from 2 nm to 4 nm height dot.
It is interesting to notice that, on the other hand,  
the width of the dot does not influence the electron-hole overlap,
so curve A (fixed $L_d$) is practically a constant over the whole $D$ range  
and the two curve (B)
and (C) corresponding to different diameter, coincide \cite{note1}. 

In the range of height values considered in Fig.3,
 the OS
varies over three orders of magnitude, so care must be taken in a
future quantum information processing experiment in order to optimize at 
the same time biexcitonic shift and OS\cite{note2}.

Let us now examine the influence of the barrier on the biexcitonic shift and 
oscillator strength.
In fig. 5 we show (upper panel, curve A) the decreasing of the biexcitonic shift 
with the barrier 
width for two 
coupled dots of 2.5 nm and 2.7 nm of height.  
We notice that it deviates from a dipole-dipole interaction behavior (curve B) and 
it is 0,5 meV 
even at 10 nm barrier.
 The reason of this behavior can be inferred from equation 
(1). The electric field
in the dot increases with the barrier width, and, as a consequence, also 
the dipole length.
The two competing effects result in a slower decreasing of the biexcitonic 
shift with the 
barrier. Curve B
corresponds to  the point-like approximation, for a fixed  dipole 
length (about 1.5 nm) corresponding to a 2 nm barrier. We see that the 
"point-like" biexcitonic shift is 
two times bigger when the barrier is 2 nm, there is a crossover 
at 4 nm, after 
which it remains smaller than the exact result. 
 The most striking consequence of the {\it barrier-dependent}
 built-in electric field, 
is that the oscillator strength strongly varies too (lower 
panel), even if the dimensions of the two dots are kept fixed.
In addition (and in contrast to Fig. 3 and 4), in this case {\it both}
OS and biexcitonic shift are {\it increased} by a smaller barrier, suggesting that the best structure for optical 
excitonic manipulations and energy-selective addressing techniques is characterized by small inter-dot barriers.
In this respect it would be very interesting to have experimental confirmation of our prediction.

In
Fig.~6 we plot the direct comparison of electron and hole distributions of 
the 2.7 nm dot, for two different barrier width (2 nm and 10 nm).
 The width of the barrier changes the internal electric 
field in the dot, that in turns strongly modifies the wavefunctions.
The dipole length changes of about 30\% 
(Fig.6), going from about 1.5 nm at 2 nm 
barrier, to 1.9  nm when the barrier is 10 nm wide.    

Compared to GaAs quantum dots, nitride dots have higher 
effective masses $m^*_{e,h}$ of both electrons and holes, lower dielectric
 constant and higher (about two 
times)
conduction and valence band discontinuities $V_{e,h}$.
 As a consequence, the GaN bulk excitonic Bohr radius is about
four times smaller than the GaAs one and 
the distributions of confined particles    
are  more "localized", because their wavelength penetration
$L_{e,h}$, where $L_{e,h}  = {\hbar\over \sqrt{2m_{e,h}V_{e,h} }}$,
depends on such parameters.

The reduction of these two characteristic lengths
 causes a faster decrease of  the electron-hole overlap,
as a function of the dipole length  
and consequently a stronger sensitivity of the oscillator
strength (roughly proportional to the electron-hole overlap in 
the  strong confinement regime  we are considering) to 
quantities such as the height of the dot (Fig. 4) or the different
built-in electric field (Fig.5), which strongly affect the dipole length.

Our results demonstrate that there exist a wide range of parameters  for 
which the
biexcitonic shift of the order of meV. This is a central prerequisite
for realizing energy-selective addressing with sub-picosecond laser 
pulses, as requested, 
for example, by all-optical
quantum information
processing schemes\cite{Biolatti1,DeRinaldis} or read-out devices\cite{damicorossi}.

\section {Blueshift and absence of molecular states}

Another interesting effect peculiar of hexagonal GaN/AlN quantum dot is 
the 
blueshift of the ground state transition when the distance between the 
dots 
is decreased,
without the lifting of the degeneracy of bonding-antibonding states.
In GaAs-based quantum dots \cite{troiani} there 
is a  redshift and an 
increasing energy
difference between bonding and
anti-bonding states, which are  spread over the whole macromolecule for 
both  electron 
and
hole; on the contrary in GaN QDs,  over the range of parameters 
here considered,
the
lowest states preserve their atomic-like shape. This depends on the fact that
 both electron and
hole effective masses and valence/conduction-band discontinuities are
much larger than in GaAs, therefore {\it decreasing} the atomic-like 
wavefunction
overlap responsible for the molecular bonding.
However, as discussed in the previous section,
 the built-in electric field is not only a function of the dot 
parameters, but 
it depends strongly on the barrier width. In particular it increases when the 
barrier is increased, to saturate at the value $F_{d} \sim
(P_{tot}^{br}-P_{tot}^{d})/(\epsilon_{0}\epsilon_{d})$, corresponding to an 
isolated dot.
For examples it increases from 5 MV/cm for a 2 nm barrier to 9 MV/cm for a 10 
nm barrier for the dot 
parameters of Fig.6 (see inset). 
In Fig.7 we consider two {\it identical}
 dots separated by increasing barrier width.
As a   consequence of the electric field change,
 the ground state energy changes of 
more than 600 meV, 
from 3337,5 meV to 2709,5 meV (upper panel of Fig.~7). Also the oscillator strength is modified 
(lower panel of fig.7) of one order 
of magnitude,decreasing  very fast up to 5 nm barrier width, showing  a 
slower decreasing afterward. 
In the particular case of {\it exactly} identical 
quantum dots, because of  the symmetry of the system, the single
exciton eigenstates are
$|{\psi}_a{\rangle}=\frac{1}{\sqrt2}(|01{\rangle}+|10{\rangle})$ 
and 
$|{\psi}_b{\rangle}=\frac{1}{\sqrt2}(|01{\rangle}-|10{\rangle})$ 
and the energy splitting is given by $2V_F$ where $V_F$
is the Foerster energy \cite{briggs}.
$V_F$ is proportional to the square of electron-hole overlap\cite{briggs}, 
therefore is much lower in GaN than in GaAs.
For example the maximum value for the parameter considered in this 
work, corresponding to 2 nm dot's height and 2 nm barrier, is only 
0.05-0.06 meV.
However we stress that when the dots are slightly different, this effect 
rapidly vanishes since it depends on the ratio $(V_F/{\Delta})^2$
where ${\Delta}$ is the energy difference between the two levels 
coupled by the Foerster transfer process. In GaN dots,
differently form GaAs, with an $8\%$ difference
in size,  ${\Delta}$ is already
of the order of hundreds of meV, 
causing this effect to be extremely difficult (if not 
impossible) to be observed experimentally.

To qualitatively investigate the effects of the different parameters 
involved we 
have calculated the ground state electron and hole distribution for some 
 GaAs-GaN "mixed case", i.e. some artificially designed molecules with 
one GaN parameter (electric field, valence/conduction band offsets, 
effective masses) substituted by the corresponding GaAs one.
We will consider a nanostructure composed by
two dots of 2.5 nm and 2.7 nm separated by a 2 nm barrier.
As a benchmark we also plot the ``pure GaAs'' case (fig.8b): 
its
ground state has 
both hole and electron distributions delocalized over the macromolecule,
forming bonding molecular states.
The corresponding ``pure GaN'' case is plotted in Fig.1 and it is characterized by  electron and hole 
distributions localized in a single dot.
The coupled GaN dots without the giant 
electric field (fig.8d ) still present an atomic character. 
The same holds for an artificial GaN 
macromolecule with GaAs effective 
masses (Fig.8c ), and an artificial GaN nanostructure with GaAs 
conduction/valence band structure (Fig.8a).
It is interesting to note that the electron and hole distribution
of a GaN dots without electric field (Fig. 8d) are more extended
than when a giant electric field is present (Fig.1). 
We conclude that the single dot confinement of the excitonic wavefunction 
in GaN QDs is a robust feature and that 
 all the parameters involved are 
responsible for the 
absence of molecular-like states in GaN coupled dots.

\section {Summary and conclusions}

We have performed a detailed investigation of the optical spectrum of 
coupled GaN quantum dots. In particular we have shown some effects 
peculiar to wurzite nitride materials compared to GaAs-based nanostructures:
the absence of ground state exciton wavefunction delocalization
even for relatively short barriers 
(up to 2 nm) due to the large 
effective mass/band offsets of nitrides; the presence of a large
 exciton-exciton 
interaction between neighbor quantum dots, caused by the giant intrinsic 
electric fields. We have also shown that
a shift of energy levels of identical 
quantum dots is expected when the distance between them is varied: such  
blueshift    
{\it is not caused by molecular coupling of the wavefunction}, but by 
the decreasing of the built in electric field as the inter-dot 
 barrier decreases. 
We have also shown how it is possible to engineer
the inter-dot biexcitonic shift and the corresponding oscillator strength
 by varying the structural parameters
(base, height, barrier) of the dots. Such analysis is crucial in the
mainframe of quantum information processing schemes\cite{DeRinaldis}
and in general for all-optical devices based on energy-selective 
 addressing\cite{damicorossi}
in which the  conditional excitonic dynamics is 
based on such quantity.

\begin{figure}
  \caption{Electron (upper panel) and hole (lower panel) particle  distribution
(dotted
line), conduction (upper panel) and
valence (lower panel) band structure (solid
line) along the growth
direction for two coupled GaN dots of, respectively, 2.5 nm and 2.7 nm of
height, separated by a 2 nm AlN barrier. }
  \label{fig1} \end{figure}

\begin{figure}
  \caption{Excitonic (solid line) and biexcitonic (dashed line)
absorption spectrum of the GaN coupled dots of fig.1.
 }
  \label{fig2} \end{figure}

\begin{figure}
  \caption{Biexcitonic shift of the
ground state transition in dot b for two
coupled GaN dots as a function of dot height and base diameter.
 In curve (A) only the  base diameter of the dots
is
changed ($D= 10 nm$);
 in curve (B) only the height of the dots
is
changed ($L_d= 2.5 nm$), while in curve (C) $D$ is varied 
proportionally to
$L_d$.
The parameters for the parabolic potential is varied in the range:
$\hbar \omega_{e} =74\div 290 meV$, $ \hbar \omega_{h} =  33\div
130 meV$. In particular the values analized in the figures are D=10nm 
($\hbar \omega_{e} =290 meV$,$ \hbar 
\omega_{h} = 130 meV$), D=11.75nm($\hbar \omega_{e} =170 meV$,$ \hbar 
\omega_{h} = 76 meV$), 
D=13.5nm($\hbar \omega_{e} =131 meV$,$ \hbar 
\omega_{h} = 59 meV$), D= 15.25nm($\hbar \omega_{e} =97 meV$,$ \hbar 
\omega_{h} = 43 meV$),  D=17nm($\hbar \omega_{e} =74 meV$,$ \hbar 
\omega_{h} = 33 meV$)}
  \label{fig3} \end{figure}

\begin{figure}
  \caption{Oscillator 
strength   of the
ground state transition in dot b 
for the same parameters of Fig.~\ref{fig3}.
Labelling as in Fig.~\ref{fig3}.
}
  \label{fig4} \end{figure}

\begin{figure}
  \caption{ Upper panel: curve (A) shows the biexcitonic shift of the
ground state transition in dot b for two
coupled GaN dots of 2.5 and 2.7 nm of height vs barrier width
between the dots.
 Curve (B) shows the corresponding
biexcitonic shift in the point-like charge 
approximation. Inset: as in main panel, but for large barrier widths.
Lower panel: oscillator strength for the same parameters described above. 
Inset: internal electric field in dot b vs barrier width.  }
  \label{fig5} \end{figure}

\begin{figure}
  \caption{Excitonic distribution in a 2.5 nm dot of a coupled dot 
nanostructure when the barrier width is 2 nm (solid line) and 10 nm (dotted 
line).}
  \label{fig6} \end{figure}

\begin{figure}
  \caption{Exciton ground state energy 
(upper panel) and oscillator strength 
(lower panel) in the dot b 
of two identical coupled GaN dots of 2.5 
nm of height 
vs barrier between the dots. }
  \label{fig7} \end{figure}

\begin{figure}
  \caption{Electron and hole particle distribution 
along the growth
direction for two coupled dots of, respectively, 2.5 nm and 2.7 nm of
height, separated by a 2.5 nm AlN barrier. 
The dot parameter sets corresponds to: 
a)GaN, except valence/conduction band offset (GaAs one);     
b)GaAs;
c)GaN, except electron/hole masses (GaAs ones);
d)GaN, except that the electric field is set to zero.}
  \label{fig8} \end{figure}

\begin{table}
\begin{tabular}{cccccc}
  	     &         & GaN &        & GaAs  &$\hspace{3em}$  \\ \hline
$E_{gap}$(eV)&{\qquad} & 3.5 &{\qquad}& 1.519 &$\hspace{3em}$\\
$m^*_e$($m_0$ units) &{\qquad}& 0.2 &{\qquad}& 0.067 &$\hspace{3em}$\\
$m^*_h$($m_0$ units) &{\qquad}& 1   &{\qquad}& 0.34  &$\hspace{3em}$\\
${\Delta}E_c$(eV)    &{\qquad}& 2   &{\qquad}& 1     &$\hspace{3em}$\\
${\Delta}E_v$(eV)    &{\qquad}& 0.5 &{\qquad}& 0.3   &$\hspace{3em}$\\
${\epsilon}$(relative) &{\qquad}&  9.6  &{\qquad}& 12.10 &$\hspace{3em}$
\end{tabular}
  \caption{Material parameters used in the calculations}
\end{table}


\begin{thebibliography}{aaaaaaaaaaaaaaaaaaaa}
\bibitem{note1} This is not a general fact, but it comes from
the constant ratio between bottom base diameter and top base diameter
assumed in the model, according to experimental data (\cite{Arley}).
This in turns causes a constant ratio between the width of electron and
hole wavefunctions and therefore a constant oscillator strength when
only the base diameter is changed.
\bibitem{note2}A figure of merit, which allows to
 optimize the interplay between biexcitonic shift and
oscillator strength,
was introduced in \cite{derinaldisapl}.
\end{thebibliography}
\end{document}